\documentclass[preprint2]{aastex}

\usepackage{lscape}

\slugcomment{Accepted by ApJ}

\shorttitle{GRB 081008}
\shortauthors{Yuan et al.}

\begin{document}

\title{GRB 081008: from burst to afterglow and the transition phase in between}

\author{F.~Yuan\altaffilmark{1}
,~P.~Schady\altaffilmark{2}
,~J.~L.~Racusin\altaffilmark{3}
,~R.~Willingale\altaffilmark{4}
,~T.~Kr\"{u}hler\altaffilmark{5,6}
,~P.~T.~O'Brien\altaffilmark{4}
,~J.~Greiner\altaffilmark{5}
,~S.~R.~Oates\altaffilmark{2}
,~E.~S.~Rykoff\altaffilmark{7}
,~F.~Aharonian\altaffilmark{8}
,~C.~W.~Akerlof\altaffilmark{1}
,~M.~C.~B.~Ashley\altaffilmark{9}
,~S.~D.~Barthelmy\altaffilmark{10}
,~R.~Filgas\altaffilmark{5}
,~H.~A.~Flewelling\altaffilmark{11}
,~N.~Gehrels\altaffilmark{10}
,~E.~G\"o\u{g}\"u\c{s}\altaffilmark{12}
,~T.~G\"{u}ver\altaffilmark{13}
,~D.~Horns\altaffilmark{8}
,~\"{U}.~K{\i}z{\i}lo\v{g}lu\altaffilmark{14}
,~H.~A.~Krimm\altaffilmark{10}
,~T.~A.~McKay\altaffilmark{1}
,~M.~E.~\"{O}zel\altaffilmark{15}
,~A.~Phillips\altaffilmark{9}
,~R.~M.~Quimby\altaffilmark{16}
,~G.~Rowell\altaffilmark{8}
,~W.~Rujopakarn\altaffilmark{13}
,~B.~E.~Schaefer\altaffilmark{17}
,~W.~T.~Vestrand\altaffilmark{18}
,~J.~C.~Wheeler\altaffilmark{19}
,~J.~Wren\altaffilmark{18}}

\altaffiltext{1}{Physics Department, University of Michigan, Ann Arbor, MI 48109}
\altaffiltext{2}{The UCL Mullard Space Science Laboratory, Holmbury St Mary, Surrey, RH5 6NT, UK}
\altaffiltext{3}{Department of Astronomy \& Astrophysics, The Pennsylvania State University, 525 Davey Lab, University Park, PA 16802}
\altaffiltext{4}{Department of Physics and Astronomy, University of Leicester, Leicester LE1 7RH, UK}
\altaffiltext{5}{Max-Planck-Institut f\"{u}r extraterrestrische Physik, Giessenbachstrasse, 85748 Garching, Germany}
\altaffiltext{6}{Universe Cluster, Technische Universit\"{a}t M\"{u}nchen, Boltzmannstrasse 2, 85748 Garching, Germany}
\altaffiltext{7}{Physics Department, University of California at Santa Barbara, 
        2233B Broida Hall, Santa Barbara, CA 93106}
\altaffiltext{8}{Max-Planck-Institut f\"{u}r Kernphysik, Saupfercheckweg 1,
        69117 Heidelberg, Germany}
\altaffiltext{9}{School of Physics,
	University of New South Wales, Sydney, NSW 2052, Australia}
\altaffiltext{10}{NASA Goddard, Greenbelt, MD 20771}
\altaffiltext{11}{Institute for Astronomy, University of Hawaii, 2680 Woodlawn Drive, Honolulu, HI, 96822}
\altaffiltext{12}{Sabanc{\i} University, Faculty of Engineering \& Natural Sciences, Orhanl{\i}$-$Tuzla 34956 {\.I}stanbul, Turkey}
\altaffiltext{13}{University of Arizona, Tucson, AZ 85721}
\altaffiltext{14}{Middle East Technical University, 06531 Ankara, Turkey}
\altaffiltext{15}{Department of Mathematics, Cag University, Tarsus 33800, Turkey}
\altaffiltext{16}{Astronomy Department, California Institute of Technology, 
        105-24, Pasadena, CA 91125}
\altaffiltext{17}{Department of Physics and Astronomy, Louisiana State
        University, Baton Rouge, LA 70803}
\altaffiltext{18}{Los Alamos National Laboratory, NIS-2 MS D436, Los Alamos, NM
        87545}
\altaffiltext{19}{Department of Astronomy, University of Texas, Austin, TX
        78712}

\begin{abstract}
We present a multi-wavelength study of GRB 081008, at redshift 1.967, by {\em Swift}, ROTSE-III and GROND. Compared to other {\em Swift} GRBs, GRB 081008 has a typical gamma-ray isotropic equivalent energy output ($\sim10^{53}$~erg) during the prompt phase, and displayed two temporally separated clusters of pulses. The early X-ray emission seen by the {\em Swift}/XRT was dominated by the softening tail of the prompt emission, producing multiple flares during and after the {\em Swift}/BAT detections. Optical observations that started shortly after the first active phase of gamma-ray emission showed two consecutive peaks. We interpret the first optical peak as the onset of the afterglow associated with the early burst activities. A second optical peak, coincident with the later gamma-ray pulses, imposes a small modification to the otherwise smooth lightcurve and thus suggests a minimal contribution from a probable internal component. We suggest the early optical variability may be from continuous energy injection into the forward shock front by later shells producing the second epoch of burst activities. These early observations thus provide a potential probe for the transition from prompt to the afterglow phase. The later lightcurve of GRB 081008 displays a smooth steepening in all optical bands and X-ray. The temporal break is consistent with being achromatic at the observed wavelengths. Our broad energy coverage shortly after the break constrains a spectral break within optical. However, the evolution of the break frequency is not observed. We discuss the plausible interpretations of this behavior. 
\end{abstract}


\section{Introduction}
The fast and precise localizations of Gamma-ray Bursts (GRBs) provided by {\em Swift}/BAT \citep{gcgmn04} in hard X-rays have enabled follow-up observations in longer wavelengths during the burst or shortly after its cessation for an increasing number of events. These observations, both from ground and from the narrow-field instruments onboard {\em Swift}, have revealed features that were not observed in the pre-{\em Swift} era. In X-rays, a canonical lightcurve \citep{nkgpg06,zfdkm06,owogp06} starts with a rapidly decaying phase that is too steep to be explained by the standard external blast wave model. This phase is often followed by a slowly decaying period, shallower than expected from the same external shock model. Only after this, the lightcurves show the normal decay and sometimes a late break, often interpreted as a jet break \citep[see e.g.][]{woogp07,lrzzb08,rlbfs09,ebpoo09}, as observed for the pre-{\em Swift} bursts. Another common but unexpected phenomena are the X-ray flares \citep{cmrfm07,fmrcm07} superimposed on the smoothly decaying segments observed for about half of the bursts. Their short time-scale variability is hard to produce externally at a large distance from the progenitor. All these features, perhaps explained by the delayed large angle internal emission \citep[e.g.][]{kpa00,lzowa06,zlz07}, continuous injection of energy \citep[e.g.][]{rm98,sm00,zfdkm06,lzz07}, delayed internal shocks or central engine activity \citep[][and references therein]{cmrfm07,fmrcm07,kgmkr09}, suggest associations with central engine properties. It is likely that the central engine is active for much longer than previously believed.

At optical wavelengths, an initial rising optical counterpart is detected for a significant fraction of bursts \citep{mvmcd07,pv08,opsdk09,kbag09,eaaab09,kgmkr09,gkmag09,gckgm09}. The temporal behaviors generally agree with predictions of the fireball forward shock model \citep{sp99, mesza06} before the onset of the shock deceleration. Alternatively, a rising afterglow can be attributed to emission in a collimated afterglow, viewed from outside the initial jet opening angle, when the shock decelerates and the relativistic beaming angle widens \citep{pv08}. A variety of lightcurve shapes, including ones with a wide peak or plateau phase, can be accommodated in this model by combination of the jet geometry and the viewing angle. Such signature of early external shock emission is not observed in X-ray, because the accompanying X-ray emission is likely obscured by the dominating internal emission. Optical observations thus provide important clues about the onset of the afterglow because emission from the external shock appears to dominate the optical emission from early times.

During the burst, optical flares apparently correlated with the hard X-ray emission are observed for several events \citep{vwwfs05,vwwag06,pwozg07,yrsrg08}. High time-resolution data plays a key role in studying the real correlation between the low and high energy emission. Such observations in optical, however, are only feasible for either a once-in-a-decade event, like the "naked eye" burst GRB 080319B \citep{rksgw08}, which was bright enough to be detected by a surveying very-wide-field optical instrument, or a bright burst lasting long enough in a usual follow-up scenario (including when the {\em Swift}/BAT is triggered on a pre-cursor). For a number of other events, an optical flare unrelated to the high energy emission is detected \citep[e.g.][]{abbbb99} or a monotonic decaying transient is observed at a flux level consistent with back-extrapolation from later observations \citep[e.g.][]{rykaa05}. These provide evidence that, at least in some cases, emission from the external shock dominates the prompt optical observations. Such diverse behaviors suggest that both internal and external shock emission may contribute to the prompt optical detections but their relative strengths vary from event to event.

\begin{figure}[h]
\epsscale{1.0}
\plotone{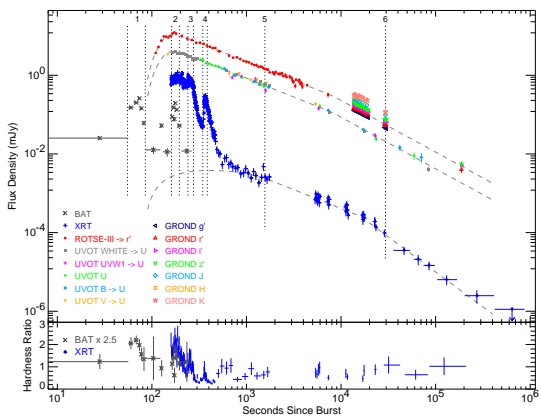}
\caption{Multi-wavelength lightcurve of GRB 081008. The UVOT flux densities are scaled to the u-band using the normalizations determined from the simultaneous temporal fit. The ROTSE-III unfiltered data are scaled to the GROND $r^\prime$-band using the normalization from the same temporal fit. The exponential to power-law decay models \citep{woogp07} for the afterglow components are over-plotted in dashed lines. Selected epochs or time intervals for spectral analysis are labeled with corresponding indices in Figure~\ref{fig2}. \label{fig1}}
\end{figure}

GRB 081008, at a redshift of 1.967, is a typical long-lasting burst detected by {\em Swift}/BAT that provides a rare opportunity to study the optical characteristics during the prompt phase. ROTSE-IIIc started imaging only 42 s after the burst trigger and before the second epoch of major $\gamma$-ray emission. An initially rising optical transient was observed, followed by two peaks, the latter coincident with a gamma-ray peak. {\em Swift} slewed immediately, and the XRT and the UVOT started observing during the second optical peak, providing well-sampled data at early times and followed the event until it dropped below the detection threshold (see Figure~\ref{fig1}). At about 13~ks after the burst, GROND started observing the afterglow in its 7 optical/IR channels. The excellent energy coverage from IR to X-ray allows us to model the spectral energy distribution with an intrinsic broken power-law, although the evolution of the spectral shape is not restricted. The overall behavior of the observed afterglow is consistent with being achromatic, while there is some hint of a slightly steeper temporal decay in the X-rays after the break.

In the following sections, we first summarize the {\em Swift}, ROTSE-III and GROND observations in $\S$ 2. We then present the multi-wavelength spectral and temporal analysis at selected epochs during the prompt and afterglow phases in $\S$ 3. We next discuss the interpretations of the observations in $\S$ 4 and finally conclude in $\S$ 5. Throughout the paper, we adopt the convention that the flux density can be described as $f_{\nu} \propto t^{-\alpha}\nu^{-\beta}$, where $\alpha$ and $\beta$ are the corresponding temporal and spectral power-law indices. Note that a negative $\alpha$ represents a rising lightcurve while a positive $\beta$, equal to the photon index minus one, corresponds to a spectrum that falls with increasing energy. We assume a standard cosmology model with the Hubble parameter H$_{\rm 0}$=70~km~s$^{-1}$ Mpc$^{-1}$ and the density parameters $\Omega_{\rm m}$=0.3, $\Omega_{\Lambda}$=0.7. All quoted errors are 1-sigma (68\% confidence), unless otherwise stated.

\section{Observations and Data Reduction}

\subsection{BAT}
{\em Swift} BAT was triggered by GRB 081008 on 19:58:09.4 UT \citep[Trigger 331093,][]{rbbbe08}. We refer to this time as the trigger time, T$_{\rm trigger}$. The BAT data were analyzed using the standard analysis software distributed within FTOOLS, version 6.5.1. The command $\it{batgrbproduct}$ was first run to provide the basic set of products. This analysis found the emission to be detected from T$_{\rm trigger}$-65.2~s and we therefore adopt this time as the onset of the burst T$_{\rm 0}$. Such a shift of the reference time only affects our analysis of the prompt and the very early afterglow, but we use it throughout the paper to be consistent.

Figure~\ref{fig1} shows the BAT flux density (at 71~keV) interpolated using the mean spectral index. The counts are binned with a minimum signal-to-noise ratio of 6 and maximum bin size of 20~s. The lightcurve displays a weak component decaying from T$_{\rm 0}$, two bright structured peaks at T$_{\rm 0}$+$\sim$65~s and T$_{\rm 0}$+$\sim$175~s and a fainter peak at T$_{\rm 0}$+$\sim$125~s. The emission drops below the BAT detection threshold at T$_{\rm 0}$+$\sim$270~s. The T90 (15-350~keV) of the burst is measured to be $185\pm39$~s. Spectral hardening and subsequent softening are observed during the first broad peak at T$_{\rm 0}$+$\sim$70~s. The average spectrum of the burst (from T$_{\rm 0}$ to T$_{\rm 0}$+263.3~s) is well fit by a power-law with exponential cutoff model. The best fit photon index and cutoff energy are 1.26(-0.32,+0.24) and 117(-50,+83)~keV (with 90\% confidence). At a redshift of 1.967 \citep{cfcb08g1,ddc08g2}, the isotropic equivalent energy release (0.1~keV-10~MeV) is $\sim$6.3$\times10^{52}$~erg, given a total fluence estimate of $\sim$6.7$\times10^{-6}$~erg~cm$^{-2}$. 

\subsection{XRT}
{\em Swift} slewed to this burst immediately and the XRT began observing in window timing (WT) mode at T$_{\rm 0}$+152.3~s. A bright uncatalogued X-ray counterpart was identified and its position distributed via a GCN alert notice. As the count rate dropped, XRT switched to photon counting (PC) mode at T$_{\rm 0}$+476~s. 

The XRT lightcurve was obtained from the {\em Swift}/XRT GRB lightcurve repository \citep{ebpto07}. The spectra were extracted from the XRT team repository, from the new products outlined in \citet{ebpoo09}. The X-ray flux densities at 2.77~keV were calculated using mean photon indices for WT and PC mode accordingly. This chosen frequency is the weighted mean of the X-ray emission assuming $\beta$=1, a typical spectral index observed, which is also similar to the afterglow spectral index of GRB 081008. The interpolations are thus not very sensitive to small changes in the photon index and the data from the two modes join smoothly. Three peaks are seen in the XRT lightcurve in Figure~\ref{fig1}. The first two have complex structures and the first peak is at a similar time as the last BAT detected pulse. The third peak is a flare superimposed on the steep decay after the second peak. After that, the lightcurve smoothly declined and steepened at T$_{\rm 0}$+$\sim$15~ks.

As observed for the first bright BAT peak, the XRT hardness ratio (1.5-10~keV/0.3-1.5~keV) curve shows a similar structure as the early lightcurve, displaying visible hardening during the flares \citep[see ][for other examples of such behavior]{brfkz05,fblck06,rmbbc06}. In general, there is also a trend of softening over time. This behavior, together with the rapid variation, indicates that the early X-ray detections are dominated by continuing burst activity, e.g. emission produced in the internal shocks. Detailed spectral analysis will be presented in $\S$ 3.

\subsection{UVOT}
UVOT began observations of the field at T$_{\rm 0}$+142.7~s. The observing sequence started with a 10~s $v$ band settling exposure and a 150~s $white$ (160-650~nm) finding chart exposure. These were followed by a 250~s $u$ band exposure, after which the filter wheel rotated through all filters, taking short 20~s exposures. One more 150~s $white$ band finding chart observation was taken at T$_{\rm 0}$+792~s. After 5~ks, longer exposures were obtained in all filters.

The optical counterpart was detected in $white$,  $v$,  $b$, $u$ and $uvw1$ bands, but not in uvm2 and uvw2 bands. This is consistent with the Lyman-$\alpha$ break at the measured redshift. In Figure~\ref{fig1}, all UVOT observed fluxes are scaled to $u$ band using the normalizations determined from the simultaneous temporal fit described in $\S$3.4. Initial analysis of the late time $white$ filter data shows the decay of the afterglow becomes very shallow after 100~ks. \citet{cfcb08g2} pointed out the presence of an extended source 2 arcsec away from the afterglow, which is within the 3 arcsec extraction radii used for UVOT. To measure the effect of this probable host galaxy, target-of-opportunity observations in the $white$ filter were requested on Nov.~13th. After removing the host contribution from the $white$ band photometry, no detections after 90~ks remain above 3 sigma. Such a correction is not attempted in other bands, given the negligible contribution from the underlying host galaxy during these earlier observations.

\subsection{ROTSE-III}
ROTSE-IIIc \citep{akmrs03}, a 45~cm robotic optical telescope located at the H.E.S.S. site in Namibia, responded to GRB 081008 automatically upon receiving the GCN notice. The first 5-seconds exposure started 107.1~s after T$_{\rm 0}$ (8.3~s after receiving the alert). An initially brightening optical counterpart was detected and reported in \citet{r08}.

ROTSE-IIIc took a total of 10$\times$5~s, 10$\times$20~s and 144$\times$60~s exposures before the target elevation dropped below the observing limit. The first ten 5-seconds exposures were taken in sub-frame mode with 3 seconds readout gap inbetween, instead of the 6 seconds full-frame readout time for later images. We took the first 70 single images and co-added the later images into sums of 5 to 10 depending on the image quality. The ROTSE-IIIc images were bias-subtracted and flat-fielded by the automated pipeline. Initial object detections were performed by SExtractor \citep{ba96}. The images were then processed with our custom RPHOT photometry program based on the DAOPHOT \citep{stetson87} PSF-fitting photometry package \citep{qryaa06}. A reference image was constructed from images taken on October 18th when the OT had faded below our detection limit.

Due to ROTSE-III's large pixel scale (3.4 arcsec), the optical counterpart is slightly blended with four surrounding objects in the ROTSE-IIIc images. To remove the possible contamination, we tried several different methods. First, we used the RPHOT functionality to subtract four PSF-scaled point sources around the OT. The lightcurve did appear to be steeper than obtained without subtraction. The two objects north of the OT are blended sources themselves but not resolved by ROTSE-IIIc. They may not be completely removed using the point source approximation, but the residual should be negligible compared to the bright OT. To confirm this, we attempted an alternative subtraction method described in \citet{ya08}. Each burst image was cross-convolved with the reference to a common PSF and the difference between the two was taken. The brightness of the OT in the difference images were then measured by matching their PSF to the stars in the convoluted reference images. The results were consistent with the RPHOT subtraction method within the uncertainties. We report here the RPHOT PSF-fitting results with the 4 contaminating sources removed. Given the above comparison, the systematic error from subtraction is smaller than the statistical errors and is not included in the reported uncertainties.

The unfiltered thinned ROTSE-III CCDs have a peak sensitivity in the $R$ band wavelength range. We thus calibrate the zero point magnitudes using the median offset from the USNO B1.0 $R$ band measurements of selected field stars. In Figure~\ref{fig1}, the ROTSE-III flux densities are scaled to match the GROND $r^\prime$-band measurements using the offsets from the simultaneous temporal fit described in $\S$3.4. 

\subsection{GROND}
The Gamma-Ray Burst Optical/NearInfrared Detector (GROND, \citealp{gbcdh08}) mounted at the 2.2~m ESO/MPI telescope at LaSilla observatory (Chile) imaged the field of GRB 081008 simultaneously in $g^\prime r^\prime i^\prime z^\prime JHK$ starting 13.6 ks after T$_{\rm 0}$, after local sunset under clear sky conditions. A total of 28 $g^\prime r^\prime i^\prime z^\prime$ images with integration times of 66~s, 115~s and 375~s were obtained during the first night post burst. At the same time, the NIR channels were operated with a constant exposure of 10~s. In addition, the field was observed with GROND at day 2, 5 and 10 after the trigger. Data reduction and relative photometry was carried out using standard IRAF tasks \citep{tody93}, similar to the procedure outlined in \citet{kgmkr09}. To exclude a possible contamination of the measured afterglow brightness with light from the host 2.2$\arcsec$ away, the results from the PSF photometry were checked against measurements taken using small apertures (0.3 and 0.5$\times$FWHM) where an aperture correction was applied. Image subtraction, using the frame taken 10 days after the burst as reference, further confirmed that the contamination is negligible. Absolute photometry was measured with respect to the primary Sloan standards SA107-1006 and SA112-805 in $g^\prime r^\prime i^\prime z^\prime$ \citep{stkrf02}, observed under photometric conditions and 2MASS field stars in $JHK$ \citep{scsws06}.

The position of the afterglow is determined as R.A. = $18^{h}39^{m}49^{s}.88$ Decl. = $-57^{\circ}25^{'}52^{"}.8$ (J2000), with a 90\% uncertainty of 0.3 arcsec. We adopt this as the best burst localization. Along the line of sight to this position, the Galactic extinction is E(B-V)=0.095 \citep{sfd98}, and the Galactic column density is $7.1\times10^{20}~{\rm cm}^{-2}$ \citep{kbhab05}.

\section{Spectral and Temporal analysis}

\subsection{Prompt Spectral Evolution}
\begin{figure}[h]
\epsscale{1.0}
\plotone{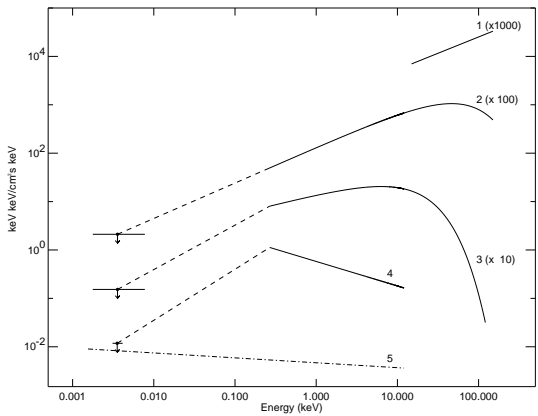}
\caption{Spectral models in the selected time intervals. Interval 1, 2 and 3 (offset for clarity) correspond to the first 3 prompt peaks. Interval 4 corresponds to the bright X-ray flare seen by the XRT and 5 is the X-ray afterglow. Optical fluxes are corrected for Galactic extinction, but are not included in the model fit in intervals 1 to 4, because the optical fluxes are likely dominated by the external forward shock and thus not related to the high energy emission. We include these points as upper limits to constrain the internal emission component in the optical. The dashed lines, connecting the data points in the X-ray and the optical regions in the same interval, do not represent the actual spectral shape. During interval 5, the emission is entirely from the afterglow and the spectrum is plotted in dashed-dot line.\label{fig2}}
\end{figure}

We first examine the spectral evolution of the burst. Available BAT and/or XRT data are extracted in three time intervals, T$_{\rm 0}$+55~s to T$_{\rm 0}$+85~s, T$_{\rm 0}$+161~s to T$_{\rm 0}$+195~s and T$_{\rm 0}$+245~s to T$_{\rm 0}$+270~s (as marked in Figure~\ref{fig1}), centered on the two broad peaks detected by BAT and the one X-ray peak shortly afterward. Because of the strong hardness ratio evolution, we limited the width of the chosen time intervals. The spectra were modeled with either a power-law or a power-law with exponential cutoff. When X-ray data were available, we included in the model a fixed Galactic absorption component and an additional variable system at the host redshift. The fit parameters for the best models are listed in Table~\ref{burstspec}. We tried modeling the spectra including a high energy power-law tail using a Band function \citep{bmfsp93}, but the power-law index above the cut-off energy is not well constrained by our data. The best fit spectral models at each epoch are plotted (with labeled offset for clarity) in Figure~\ref{fig2} as solid lines.

\begin{deluxetable}{cccccccc}

\tablecaption{Best-fit spectral parameters to the burst and the flares \label{burstspec}}
\tabletypesize{\scriptsize}

\setlength{\tabcolsep}{0.18in}

\tablehead{\colhead{Interval} & \colhead{T$_{\rm range}$-T$_{\rm 0}$} & \colhead{Host nH} & \colhead{$\beta$} & \colhead{$E_{cut}$} & \colhead{$\chi^2/dof$} & \colhead{Instrument} \\ 
\colhead{} & \colhead{(s)} & \colhead{($10^{22}~{\rm cm}^{-2}$)} & \colhead{} & \colhead{(keV)} & \colhead{} & \colhead{} } 

\startdata
1 & 55 - 85 & \nodata & $0.33\pm0.07$ & \nodata & 58/56 & BAT  \\
2 & 161 - 195 & $1.06_{-0.27}^{+0.32}$ & $0.26_{-0.05}^{+0.04}$ & $63.2_{-9.8}^{+13.3}$ & 201/184 & BAT \& XRT \\
3 & 240 - 275 & $1.01_{-0.26}^{+0.29}$ & $0.58_{-0.14}^{+0.12}$ & $14.9_{-5.3}^{+8.5}$ & 200/194 & BAT \& XRT \\
4 & 345 - 385 & $1.77_{-0.32}^{+0.38}$ & $1.51_{-0.12}^{+0.13}$ & \nodata & 97/75 & XRT \\
\enddata

\tablecomments{All errors quoted correspond to 90\% confidence intervals.}

\end{deluxetable}

In both interval 2 and 3, the high energy spectra are well described by a power-law with exponential cutoff model where the cutoff energy drops from above 60~keV in interval 2, to 15~keV in interval 3, and the photon index softens. The measured local column density (1.06(-0.27,+0.32)$\times10^{22}~{\rm cm}^{-2}$ and 1.01(-0.26,+0.29)$\times10^{22}~{\rm cm}^{-2}$) shows good agreement between intervals. During interval 1, the BAT data are fit by a power-law model with a relatively hard photon index ($\Gamma\sim$1.33). Given the general trend of softening, it is likely that the cutoff energy at this early epoch is close to or above 150~keV and thus not well constrained by the BAT energy coverage. \citet{ssbbc09} has demonstrated that a reasonable estimate of the peak energy (in ${\nu}f_{\nu}$ space) can be achieved using the power-law photon index in the BAT energy range if the latter is between 1.3 and 2.3. Our value of 1.33 for the 1st interval is just inside this range and corresponds to a peak energy of 196(-47,+587)~keV, assuming a cutoff power-law model as the actual spectral form. For the later two intervals, the peak energies can be explicitly calculated as $(1-\beta)E_{cut}$, which are $\sim$47~keV and $\sim$6~keV respectively. These numbers agree with the estimates following the method in \citet{ssbbc09}, although the BAT-alone power-law photon index in the 3rd interval is not well constrained and exceeds 2.3.

During the 2nd and 3rd intervals, optical data are also available, but they are not included in the spectral fit for two reasons. First, emission from a different origin (e.g. external forward shock) is likely dominating the optical data (see $\S$3.3 and $\S$3.4). Second, only data in the $white$ filter (or unfiltered) is available, thus any optical deficit below the extrapolation from high energy observations can be compensated by an arbitrary value of local extinction, while any optical excess is hard to accommodate within the chosen models. Instead, we plot the $white$ band fluxes (corrected for Galactic extinction and plotted with the same offsets as the corresponding high energy spectra)  in Figure~\ref{fig2} and connect them directly to their high energy counterparts with dashed lines, just to show how they compare. The optical detections are considered as upper limits to constrain the internal emission component. If internal emission has contributed to at least part of the optical observations, they do not exceed the extrapolations from XRT and BAT detections at any time.

\subsection{X-ray Flare}
A bright single X-ray flare is superimposed on the rapid decay phase after the second broad X-ray peak. It has a similar hardness ratio evolution as the earlier peaks. It also shows a fast-rise-exponential-decay (FRED) profile that is typical of a GRB pulse. Its rapid rise and decay supports an identification with extended central engine activity. We thus analyze the flare within the same frame as the prompt emission.

We extract the X-ray spectrum from T$_{\rm 0}$+345~s to T$_{\rm 0}$+385~s (noted as interval 4), centering on the flare. UVOT $v$ band and ROTSE-III detections are available during this period, but there is no sign of a flare in the optical indicating that the emission is dominated by the afterglow. So we fit the XRT data alone. The spectrum is reasonably fit by a single power-law model with a photon index of 2.5 (see Table~\ref{burstspec}), which is much softer than the prompt emission and agrees with the softening trend. The inferred column density is somewhat higher than those from the earlier times, probably caused by curvature in the real spectral shape. This would suggest that $E_{peak}$ is near or within the XRT energy range at this time.

\subsection{Early Optical Afterglow}
\begin{figure}[h]
\epsscale{1.0}
\plotone{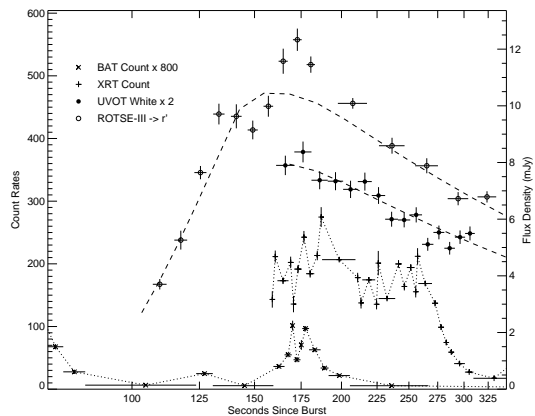}
\caption{Contemporaneous BAT, XRT and optical (ROTSE-III and UVOT $white$) observations. Time bins for the BAT and XRT data are determined by Bayesian block analysis (see text in $\S$3.3). The high energy data points are connected by dotted lines as guidance to the eye. The UVOT event data is binned in 10s intervals. The smooth rise and decay model is plotted over the optical to show that no significant deviation is detected. \label{fig3}}
\end{figure}

The ROTSE-III observations show an initially rising optical transient with two peaks at T$_{\rm 0}+\sim$135~s and T$_{\rm 0}+\sim$170~s (see Figure~\ref{fig1} and \ref{fig3}). The later one is almost simultaneous with the BAT and XRT peaks and may be attributed to a related prompt optical flare. The first one, however, does not appear to be correlated with the high energy emission. As shown in $\S$3.4, the overall optical lightcurve is well-fit by one dominating component. We therefore interpret this first peak in terms of external shock emission. A detailed discussion of this interpretation will be presented in $\S$4.1.

\begin{deluxetable}{cccccc}
\tablecaption{Afterglow temporal best-fit parameters \label{tempfit}}
\tabletypesize{\scriptsize}
\setlength{\tabcolsep}{0.18in}

\tablehead{\colhead{Energy Range} & \colhead{Rise time} & \colhead{Start of Decay}  & \colhead{$\alpha_{1}$} & \colhead{T$_{\rm break}$} & \colhead{$\alpha_{2}$} \\
  \colhead{} & \colhead{$t_a$(s)} & \colhead{$T_a$(s)} & \colhead{}& \colhead{(s)} & \colhead{} } 

\startdata
X-ray      & 130\tablenotemark{*} & 1800\tablenotemark{**} & $0.96\pm0.10$  & $15870\pm8623$ & $1.78\pm0.25$   \\
Optical/IR & 130\tablenotemark{*} & $136\pm18$ & $0.96\pm0.03$  & $7834\pm2160$  & $1.313\pm0.048$ \\
X-ray \& Optical/IR \tablenotemark{***} & \nodata & \nodata & $0.81\pm0.07$  & $15139\pm8150$ & $1.71\pm0.13$   \\
\enddata

\tablenotetext{*}{Fixed parameters, to a value estimated from the optical fit.}
\tablenotetext{**}{The plateau is hidden by the prompt emission, thus no error is estimated.}
\tablenotetext{***}{Only data after T$_{\rm 0}$+750s are fit by a smoothly joined broken power-law model, with a smooth factor, s=1.}
\tablecomments{All errors quoted correspond to 90\% confidence intervals.}

\end{deluxetable}

Figure~\ref{fig3} shows a close look at the BAT, XRT and optical lightcurves before T$_{\rm 0}$+345~s. The BAT and XRT data are binned in intervals determined by Bayesian block analysis \citep{s98}. For BAT, this analysis is carried out by $\it{battblock}$ on the 1~s-binned lightcurve data. For XRT, only WT data are used and the background ($\sim$1~ct/s) is negligible compared to the bright signal. We thus adapted the algorithm used in $\it{battblock}$ for event files and consider only the change points resulting in time bins longer than half a second to ensure comparable signal to noise ratios. The high energy lightcurves show significant variations on a similar time-scale; while no convincing short-time-scale variability is detected in optical. This is not surprising though, as the optical curve connects smoothly into a monotonic decline afterward.

\subsection{Multi-wavelength Afterglow}
We adopt the two component exponential to power-law decay model in \citet[][Eq~(2) and descriptions within]{woogp07} to characterize the temporal flux evolution in X-ray and optical. In this model, the first component corresponds to the prompt emission, and the second relates to the afterglow emission. All optical observations are sufficiently fit by one afterglow component. In the X-rays, the second component is not revealed until the end of the last flare. So the rise time ($t_a$) of this second component is fixed to 130~s according to the optical fit. The best fit models for the afterglow components are over-plotted in Figure~\ref{fig1} and the parameters are tabulated in Table~\ref{tempfit}.

For both optical and X-ray, a later temporal break is required in the model. The break times in the two energy ranges (15870$\pm$8623~s and 7834$\pm$2160~s) are consistent at 90\% confidence, but the final decay ($\alpha_2$ in Table~\ref{tempfit}) in the X-ray is significantly faster than in the optical. We further attempted fitting all energy bands after T$_{\rm 0}$+750~s simultaneously with a smoothly joined broken power-law model (f(t)=$a[(t/t_b)^{-s\alpha_1}+(t/t_b)^{-s\alpha_2}]^{1/s}$), with a small smoothness factor, s$\sim$1. Although the overall fit is acceptable, giving a reduced Chi-squared of 1.17 for 239 degrees of freedom (see Table~\ref{tempfit} for the best fit parameters), the fit parameters are dominantly determined by the well-confined optical data. Some systematic deviations from the model, not reflected in the Chi-squared statistics, are noticed in the X-rays. Between T$_{\rm 0}$+7~ks and T$_{\rm 0}$+28~ks, the X-ray points are mostly above the model, and after T$_{\rm 0}$+30~ks, the X-ray data are all below the model. Our data are thus consistent with displaying a smooth achromatic temporal break, but a chromatic behavior is not confidently ruled out in case of a subtle evolution.

We then examine the spectral energy distributions (SEDs) in selected afterglow epochs. In the X-rays, no significant spectral or hardness ratio changes can be detected. In the optical, the UVOT detections before the break and the GROND observations after the break are poorly fit by one power-law spectral index plus the same amount of local extinction. This is caused either by spectral evolution or a spectral break at the red (blue) end of the UVOT (GROND) coverage, at around 2.5~eV. This latter conjecture is confirmed by further analysis.
 
\begin{deluxetable}{ccccccccc}
\tablecaption{Best-fit afterglow spectral parameters \label{afterglowspec}}

\tabletypesize{\scriptsize}
\setlength{\tabcolsep}{0.03in} 
\tablehead{\colhead{T-T$_{\rm 0}$} & \colhead{Energy Range} & \colhead{Host Type} & \colhead{Host Extinction} & \colhead{nH} & \colhead{$\beta_{opt}$} & \colhead{$E_{break}$} & \colhead{$\beta_{x}$}& \colhead{$\chi^2/dof$} \\ 
\colhead{(ks)} & \colhead{} & \colhead{} & \colhead{E(B-V)} & \colhead{($10^{22}~{\rm cm}^{-2}$)} & \colhead{} & \colhead{(eV)}& \colhead{} & \colhead{}} 

\startdata
1.5 & XRT only & \nodata & \nodata & $0.47_{-0.33}^{+0.36}$ & \nodata & \nodata  & $1.06_{-0.13}^{+0.14}$ & 46/43  \\
1.5 & XRT \& UVOT & MW & $0.15_{-0.03}^{+0.04}$ & $0.62_{-0.23}^{+0.25}$ & $1.14\pm0.05$ & \nodata & $1.14\pm0.05$ & 52/46  \\
1.5 & XRT \& UVOT & LMC & $0.11_{-0.03}^{+0.02}$ & $0.58_{-0.22}^{+0.25}$ & $1.12\pm0.04$ & \nodata & $1.12\pm0.04$ & 53/46  \\
1.5 & XRT \& UVOT & SMC & $0.08\pm0.02$ & $0.54_{-0.22}^{+0.24}$ & $1.10\pm0.04$ & \nodata & $1.10\pm0.04$ & 55/46  \\
29.5 & XRT, UVOT \& GROND & MW & $0.18_{-0.07}^{+0.09}$ & $0.97_{-0.60}^{+0.85}$ & $-0.13_{-0.41}^{+0.30}$ & $2.4\pm0.2$ & $1.12_{-0.08}^{+0.11}$ & 11/14 \\
1.5 \& 29.5 & XRT, UVOT \& GROND & MW & $0.16_{-0.03}^{+0.04}$ & $0.65_{-0.21}^{+0.24}$ & $-0.06_{-0.18}^{+0.17}$ & $2.4_{-0.1}^{+0.2}$ & $1.14\pm0.05$ & 72/63 \\
\enddata

\tablecomments{All errors quoted correspond to 90\% confidence intervals.}

\end{deluxetable}

The overall shapes of the SEDs are modeled from optical/IR to X-ray with all available data before and after the break. Assuming they belong to the same synchrotron emission component, we allow either one or no spectral break within the energy range. Before the break, we extract an X-ray spectrum in the time interval from T$_{\rm 0}$+765~s to T$_{\rm 0}$+1778~s (first orbit) and from T$_{\rm 0}$+5286~s to T$_{\rm 0}$+7562~s (second orbit) to maximize the signal to noise ratio, given that no significant spectral or hardness ratio changes is detected in the X-rays. This spectrum is scaled to the X-ray count rate at T$_{\rm 0}$+1.5~ks to minimize the interpolation of the UVOT data. The UVOT mean count rates at T$_{\rm 0}$+1.5~ks are estimated by fitting the nearby data points (within 500~s on each side) in each band with power-laws. The errors are estimated using a bootstrap approach, where each data point is randomly re-sampled from a Gaussian distribution centered on the observed count rate with a sigma equal to the measured error. The UVOT count rates are converted to XSPEC compatible spectral files with the corresponding response matrices downloaded from the HEASARC Calibration Database (version 20041120v104). 

The SED can be fit by a single power-law model with fixed Galactic dust and gas absorption and additional dust and gas absorption at the host redshift (see Table~\ref{afterglowspec}). This model is supported by the well-matched initial temporal power-law decay indices ($\alpha_1\sim$0.96) in the optical and the X-ray. The relatively small amount of host extinction required and the lack of coverage at the rest-frame 2175$\mbox{\AA}$ Galactic absorption feature do not allow us to distinguish between the Milky Way (MW), Large Magellanic Cloud (LMC) or Small Magellanic Cloud (SMC) environment. 

After the break, another X-ray spectrum is constructed between T$_{\rm 0}$+16~ks and the last detection at T$_{\rm 0}$+400~ks. This spectrum alone has a relatively low signal to noise. We scale this X-ray spectrum to the latest epoch where the afterglow was detected in the $J$, $H$ and $K$ bands (at T$_{\rm 0}$+$\sim$29.5~ks) and convert the contemporaneous GROND detections to XSPEC compatible files. The UVOT $u$ and $b$ band rates are also interpolated to this epoch by fitting the neighboring two detections (about 30~ks apart) with power-laws as described above.

A single power law model does not produce an acceptable fit to the overall SED, regardless of the choice of local dust and gas absorption. At least one spectral break is required between X-ray and optical. Our best fit model (see Table~\ref{afterglowspec}) constrains a break at $\sim$2.4~eV, in the UVOT v-band. Such a low break frequency is not well covered by the earlier UVOT data, and the two SEDs thus have consistent shapes. A MW extinction curve is used because it provides a best fit and is marginally preferred during the earlier epoch. Nevertheless, we do not consider this as a conclusive property of the environment because of the possible uncertainty in our selected intrinsic spectral shape, e.g. the smoothness of the break is not considered in our model.

\begin{figure}[h]
\epsscale{1.0}
\plotone{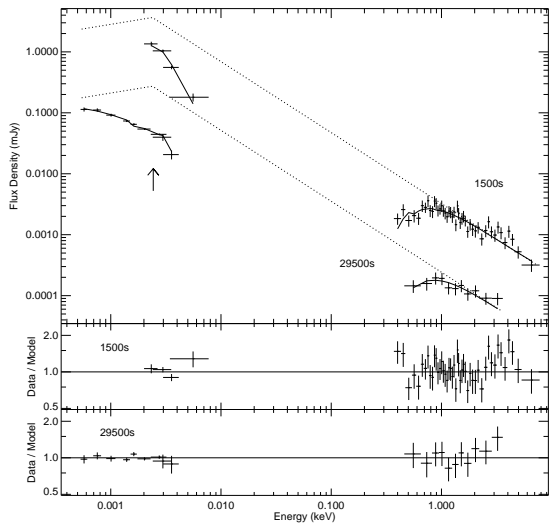}
\caption{Combined XRT, UVOT and GROND spectral energy analysis during the afterglow (epoch 5 and 6). The local extinction and absorption properties are tied. The fit parameters are listed in Table~\ref{afterglowspec}. The absorbed models are over-plotted in the upper panel. Dotted lines show the intrinsic power-law and broken-power-law components. The location of a spectral break is indicated by an upward arrow. The residual ratios are plotted in the bottom panels. \label{fig4}}
\end{figure}

Figure~\ref{fig4} shows a simultaneous fit of the two SEDs with a broken power-law model with Galactic dust and gas absorption and dust and gas absorption at the host redshift. We tie the dust (following a MW extinction law) and gas properties for a consistent environment throughout the afterglow phase. The fit is fairly good except in the $uvw1$ band. A plausible explanation for the under-estimate in $uvw1$ is that our model applies a sharp cutoff above the Lyman-break which might not be true. The best fit model parameters are consistent within errors with the estimates obtained in narrower energy bands and smaller time coverages as described above. Also, the column density derived is consistent with the estimate during prompt interval 2 and 3 (see Table~\ref{burstspec}).

In summary,  the observed optical and X-ray afterglow emission started off following similar temporal decay rates ($\alpha_{1}\sim$0.96) and displayed smooth temporal breaks at similar times (at $\sim$15~ks). While the two SEDs before and after the break were characterized intrinsically by a power-law model and a broken power-law model respectively, they had consistent shapes because the low break frequency was not well covered during the first epoch. Besides, the optical/IR observations by GROND between T$_{\rm 0}$+13~ks and T$_{\rm 0}$+30~ks do not show evidence of a moving spectral break. The overall afterglow behavior was consistent with being achromatic. Nonetheless, there is some hint of a slightly steeper decay in the X-rays after the break. Because the afterglow was not observed in IR and near-IR at early times, probable evolution of the spectral break is not constrained, e.g. it could have increased over time from below optical to the observed 2.4~eV.

\section{Discussion}

\subsection{Early Optical Afterglow}

\subsubsection{Reverse or Forward Shock}
For the initial optical observations, fixing the beginning of the external shock to the burst time T$_{\rm 0}$ yields a temporal rise index of $\alpha = -5.0\pm0.6$. This is too rapid to be considered as the pre-deceleration phase of the forward shock \citep[$t^2$ or $t^3$ in ISM, slower than $t^{1/2}$ in Wind medium, see][]{pv08}, but agrees well with the prediction for a reverse shock rising in a thin shell case when the burst duration is less than the deceleration time \citep{k00,zkm03}. Back-extrapolation from late time afterglow observation shows that the reverse shock and forward shock must have comparable peak luminosities. The analysis of the later afterglow emission suggests the forward shock characteristic synchrotron frequency ($\nu_m$) may have passed below the optical by T$_{\rm 0}$+200~s. If the micro-physical parameters are comparable in the reverse and forward shock regions, the reverse shock should have a much lower characteristic synchrotron frequency \citep{sp99} and thus have a very weak optical emission. The detection of reverse shock emission would then point to a more magnetized reverse shock region \citep{zkm03,kp03,jf07}.

However, there are several questions to be asked for this interpretation. First, is the thin shell assumption correct? Although the overall prompt BAT and XRT detections extend to after the beginning of the optical detections, the second epoch of strong burst activity emerged once the initial rapid rise in the optical had stopped. The temporal and spectral similarity between the flares indicate that the second epoch of high energy emission was from the same emitting region as the earlier bursts of emission, i.e. closer to the central engine than external medium. This later part of the ejecta would then run into the external medium after the initial shock has been established. We thus can ignore the effect of this later injected kinetic energy when considering the early rising lightcurve. Suppose the bulk of kinetic energy was injected into the external medium by the materials producing the first flare ending at T$_{\rm 0}$+100~s, the shock can be reasonably considered impulsive.

This leads to the second question: whether the onset of the afterglow is chosen correctly as T$_{\rm 0}$? \citet{qryaa06} have pointed out that different choices of time origin may lead to different interpretations of the early data \citep[also see ][for a discussion in the X-rays]{tgcmc05}. \citet{lb06} explored how varying central engine activity would affect the self-similar evolution in the early afterglow and its lightcurve. Their simulation is limited to a power-law energy release, but gives a good schematic picture. They found that if the burst energy release is concentrated toward the end of the burst duration, the self-similar evolution would deviate from one simple power-law behavior. For GRB 081008, the emission at T$_{\rm 0}$+65~s is about one order of magnitude higher than at T$_{\rm 0}$, it is therefore proper to choose a delayed time origin in a power-law fit. Such a small time shift has a negligible effect on the temporal index inferred for the later smooth decay. But if we choose the reference time to be T$_{\rm 0}$+45~s, the initial rise phase would have a temporal index of $-3.1\pm0.4$, fully consistent with the pre-deceleration of the forward shock. For any choice of reference between T$_{\rm 0}$+25~s and T$_{\rm 0}$+62~s (including the beginning of the first big pulse detected by BAT), the estimated temporal power-law index is consistent with -3 within 2~sigma. Given the earlier discussion about the reverse shock and the fact that the lightcurve is well fit by one smooth rise and fall component, we prefer the external forward shock interpretation for this rising lightcurve observed by ROTSE-III.

Assuming the first emission peak in the optical marks the onset of the forward shock deceleration, we can apply the adjusted rise time (T$_{\rm 0}$+135~s~-~T$_{\rm 0}$+45~s) of 90~s and the energy released before T$_{\rm 0}$+135~s (see $\S$4.1.2) into Eq (1) in \citet{mvmcd07} to obtain a Lorentz factor of $\Gamma_{\rm dec}\sim$225 at the deceleration radius and an initial Lorentz factor $\Gamma_{\rm 0}$ of $\sim$450 \citep[$\sim2\Gamma_{\rm dec}$;][]{mesza06}. We have assumed a particle density n=1~${\rm cm}^{-3}$ and a radiative efficiency $\mu$=0.2. The deceleration radius is then estimated as ${\sim}2{\Gamma}^{2}ct{\sim}3\times10^{17}$~cm.

\subsubsection{Refreshed Shock?}
After the forward shock start decelerating, the passing of the synchrotron frequency and/or cooling frequency across the optical band may produce temporal breaks. But a second peak, as observed for GRB 081008, is hard to explain in a model involving only one external forward shock component. The temporal coincidence with a high energy flare suggests possible contribution from internal emission. This scenario is hard to rule out given the slight optical excess ($\sim$4~sigma) above the afterglow model seen in Figure~\ref{fig3}. However, the lightcurve also shows a probable dip below the smooth, one component model, and the decay after the second peak joins smoothly into the later phase. We thus explore other interpretations for the double peaked lightcurve.

In a reasonable contemplation in the frame of the fireball model, the shells of material producing the late burst activity would reach the external medium and catch up with the early decelerated shells at some moment and re-energize the forward shock. This later scenario results in an irregular ``step'' structure, as additional energy is injected over a time period comparable or longer than the second epoch of burst activity. It has been noted that refreshed shocks from slow or late shells may alter the shape of the afterglow lightcurves \citep{rm98,kp00}. It is, however, hard to directly relate such structure with the internal emission as they are produced by different mechanisms at different regions. Here, we provide a simplest possible test on this speculation.

If a second batch of kinetic energy is injected into the forward shock, a new self-similar solution would be established sometime after the cessation of all burst activity. Indeed, we didn't observe any apparent deviation from a power-law decay after T$_{\rm 0}$+200~s. We try to estimate the synchrotron emission before and after the refreshed shocks. Without the additional energy injection, we can assume the afterglow would decay monotonically after the first peak at the same rate as observed in the later afterglow. At a time after T$_{\rm 0}$+200~s, the observed flux is 50-60\% higher than this extrapolation, including uncertainties from the flux measurement at the first peak and the normalization of the later decaying lightcurve.

We then estimate the energy release in the prompt emission before and after T$_{\rm 0}$+135~s (the first optical peak). We estimate the fluence between 0.1~keV and 10~MeV in the rest frame. In the observer's frame, the 0.1~keV lower bound fits nicely within the XRT sensitivity range. Before T$_{\rm 0}$+135~s, only BAT observations are available and the average spectrum can be fit by a power-law with exponential cutoff model. The cutoff energy at 157~keV, although above the BAT energy threshold, agrees well with calculation using the power-law photon index following \citet{ssbbc09}. We estimate the fluence during this period to be $\sim4.7\times10^{-6}$~erg~cm$^{-2}$. The hard X-ray emission dropped below the BAT detection threshold at T$_{\rm 0}$+$\sim$265~s, but it is likely that the central engine activity continued with softer emission detected mainly in soft X-rays. Between T$_{\rm 0}$+155~s and T$_{\rm 0}$+210~s, simultaneous XRT and BAT detections constrain the spectral shape well, and using a power-law model with an exponential cutoff, we estimate a total fluence of $\sim2.5\times10^{-6}$~erg~cm$^{-2}$. After this, the peak energy shifts further toward lower energies and the fluence in the next 100s is only one fourth of the previous interval (with cutoff energy at $\sim$10~keV). The additional energy release between T$_{\rm 0}$+135~s and T$_{\rm 0}$+155~s is even smaller. From T$_{\rm 0}$+135~s to T$_{\rm 0}$+$\sim$310~s, we estimate a total prompt fluence of $\sim3.4\times10^{-6}$~erg~cm$^{-2}$. This is about 70\% of the total energy released before T$_{\rm 0}$+135~s, and consistent with the flux increase calculated within the large uncertainties. Note that the overall energy output is about 20\% higher than the estimate in $\S$2.1 using an average gamma-ray spectrum and the BAT detected burst duration.

The X-ray flare at T$_{\rm 0}$+$\sim$400~s is likely generated by the same internal mechanism. We can estimate its emission in the same energy range as above. The spectrum is so soft (with photon index greater than 2) that we can directly integrate it in the high energy range. No extrapolation below the XRT energy is necessary with our choice of energy range, so we do not have to worry about a spectral break below 0.3~keV. The ``bolometric'' fluence calculated is two orders of magnitude lower than the previous prompt emission. Therefore, if the contribution to the forward shock is proportional to the prompt luminosity, the change in the forward shock emission is comparable to the error bars in the optical measurements and too small to be noticed. 

A caveat for this calculation is the assumption that the relative radiative efficiency in the internal and external processes is constant although the relationship between the two is not straightforward. During the burst, the peak emission frequency down-shifted from $\gamma$-ray to X-ray. The X-ray emission could be from slower or delayed ejecta, or alternatively from collisions with a smaller Lorentz factor dispersion and thus lower efficiency in converting the kinetic energy. At the forward shock front, the conversion efficiency may depend on the condition of the previously shocked medium and therefore requires a detailed simulation to determine. Another overlooked factor is the evolution of the synchrotron spectral shape, as the flux is estimated in a narrow optical band.

\citet{vwwag06} showed that the early optical lightcurve of GRB 050820A can be modeled with a prompt component tracing the $\gamma$-ray emission and two afterglow components. The afterglow components, correspondingly produced by the precursor and the main event ejecta, have consistent ratios between the prompt fluence and afterglow amplitude. This agrees with our finding although their first afterglow component is faint and not well-sampled for a detailed study. A similar double peaked prompt lightcurve is observed for GRB 061007 \citep{eaaab09}. The overlapping high energy pulses indicate continuous energy input into the external shock and thus modeling of the early lightcurve in a thick shell case may be more appropriate. For that event, the initial very rapid rise of the optical emission is also problematic in the current forward shock model.

\subsection{Late Afterglow Evolution}
Before the temporal break and above $\sim$2.4~eV, optical and X-ray appear on the same segment of a synchrotron spectrum, we thus use the method described in \citet{rlbfs09} to find all scenarios consistent with the observed temporal and spectral indices. The temporal decay ($\alpha$=0.96$\pm$0.03) marginally agrees with the value expected ((3$\beta$-1)/2$\sim$1.2, for $\beta\sim$1.1) for above the cooling and characteristic synchrotron frequencies, regardless of slow or fast cooling regime and the external medium density profile. In such a scenario, both the cooling frequency and the characteristic frequency are below or close to the optical by T$_{\rm 0}$+200~s.
 
For all the other closure relations considered, the temporal decay is shallower than predicted if no energy injection is present. In terms of the canonical X-ray lightcurve \citep{nkgpg06,zfdkm06,owogp06}, this decay index is within the range observed for the shallow decay phase (phase II). This phase, detected in a significant fraction of the bursts, is generally explained as due to prolonged energy injection. However, the source of energy is unclear as well as the mechanism for fine tuning the input energy into the forward shock smoothly over such long period of time.

We apply the evolution constraints in \citet{rlbfs09} to the X-ray data and find the cause of the break is consistent with either a jet geometry or the cessation of energy injection. For both types of break, achromatic behavior extending to the optical range is expected, consistent with our observations.

After the temporal break, our data tentatively suggest a spectral break around 2.4~eV, although it is hard to constrain the exact spectral shape and the environment properties at the same time. Such a break is hard to be accommodated in the standard fireball model. A synchrotron spectrum is characterized by three possible break frequencies, a cooling frequency ($\nu_c$), a characteristic frequency ($\nu_m$) and a self-absorption frequency ($\nu_a$) \citep[e.g.][]{sp99,gs02}. All three frequencies should evolve over time. The observed break at 2.4~eV is too high to be a self-absorption frequency. Given a spectral index below the break (-0.13\{-0.41,+0.30\}) consistent with -1/3, the shape of the spectrum can be expected in a slowing cooling regime ($\nu_m<\nu_c$) around $\nu_m$ \citep{gs02}. However, $\nu_m$ should always decease over time as the forward shock decelerates, unless the shock is reenergized by significant late-time injection, which does not seem to be the case here. We do not observe the expected rise or flat evolution in IR (below the break) and do not detect a higher break frequency in our earlier SED. Finally, our data do not rule out an increasing break frequency, as expected for a cooling break in a wind medium \citep{cl00}, but the spectral index difference across the break ($\sim$1) is much larger than the predicted amount of 0.5.

If some chromatic effect exists around the temporal break, the passage of a cooling frequency in a wind medium produces the correct trend of a break moving from optical to X-ray. The expected temporal index change of 0.25 is similar to the observed value in the optical ($\sim$0.3) but not enough to account for the amount of change detected in the X-rays ($\sim$0.8). In any case, the temporal break at $T_{\rm 0}$+15~ks is not likely to be caused by the movement of the observed spectral break, because there is no evidence that the break frequency has yet reached the X-ray regime by $T_{\rm 0}$+29.5ks, as would be expected in such a case.

In short, the observed temporal break is consistent with being a jet break or the cessation of energy injection. Before the break, the multi-wavelength observations are best interpreted by the cooling and characteristic synchrotron frequency being below the optical. No satisfactory explanation is found for the later spectral shape and its probable evolution. It is viable to assume that the spectral break has increased over time, as the cooling frequency in a wind medium, but such evolution is not likely to be responsible for the observed temporal break.

\subsection{Burst Energetics}
If the temporal break at $T_{\rm 0}$+$\sim$15~ks is interpreted as a jet break, the jet opening angle can be estimated as $2.1^\circ$ or $2.4^\circ$ for ISM \citep{sph99} or a wind medium \citep{bfk03} respectively. In the calculations, we have assumed a particle density n=1${\rm cm}^{-3}$, a characteristic wind density A$_{*}$=1, a radiative efficiency $\mu$=0.2 and a kinetic energy $E_{k}=E_{\rm iso}/\mu$. The collimation corrected total gamma-ray energy release (E$_\gamma$) is hence $4.2\times10^{49}$~erg and $5.5\times10^{49}$~erg for ISM and wind medium respectively. This number would be slightly larger if early X-ray detections are also considered (see $\S$4.1.2).

For a jet break, the change of the temporal index ($\sim$0.8) favors an ISM-like medium \citep{pana05}.  Both the inferred opening angle and the collimated energy output are on the small side but well within the distribution for {\em Swift} detected GRBs \citep[e.g.][]{lrzzb08,rlbfs09}. The nature of the break is, however, uncertain given the previous discussions on a probable secondary chromatic effect. A later jet break, occurring close to or shortly after the afterglow dropped below the detection threshold of all our instruments ($\sim$400~ks), would also give a sensible estimate of the opening angle and E$_\gamma$. 

\section{Conclusions}
With energetics typical among {\em Swift} detected events, GRB 081008 happens to last long enough to have overlapping phases with the early follow-up observations in X-ray and optical. It is well behaved in the sense that all early canonical phases in X-ray and optical were detected. A key feature of the high energy emission is the hardness ratio evolution in accordance with the flux fluctuation throughout the burst and the X-ray flares. The temporal continuity and spectral softening suggest that early X-ray observations originate from extended internal activities, but either with a gradually decreasing efficiency or a declining energy output from the central engine.

In the optical, the detections seem to be dominated by afterglow emission at all times. An intriguing question to ask is what happens to the material producing the late burst emission. In the frame of the fireball model, they will run into the ambient medium and refresh the external shock. Whether this signature is observed for GRB 081008 is uncertain. If such early refreshed shocks can be observed, their relative strength to the prompt emission provide important clues to the burst mechanism and production of the afterglow. Refreshed shocks have been used to explain late-time lightcurve bumps \citep{gnp03,jbg06}. In those cases, slow shells catch up with the forward shock front hours to days after the initial burst. The existence and properties of the slow shells are only inferred from the lightcurve shapes. Observations directly linking the ejecta material and refreshed shocks are achievable for long bursts with discrete episodes of activities or for events with very bright X-ray flares if they are produced in a similar way as the prompt burst emission. The latter cases may produce cleaner chasers of the refreshed shock with less contamination from internal optical emission \citep[however, see][]{kgmkr09}. \citet{fblck06} have attempted to relate the subsequent bumps in the X-ray lightcurve with the bright X-ray flare in GRB 050502B. No solid evidence of refreshed shocks associated with flares has been observed in the optical so far, probably due to limited time coverage of the data or incomplete understanding of the modification to the afterglow lightcurve.

The importance of well-sampled multi-wavelength data has been demonstrated. In general, the observed afterglow emission is consistent with displaying an achromatic temporal break. Such a break can be interpreted as due to a collimated outflow (jet) or the end of energy injection. However, there is a hint of a steeper decay in the X-rays after the break. While for GRB 081008, the deviation is rather subtle, similar steeper decay in the X-rays than in the optical has been observed for some other events \citep[e.g.][]{rmysa06,pwbdh09}. Using data from GROND, UVOT and XRT, we are able to model the SED of the afterglow from IR to X-ray and constrain a probable spectral break within the optical regime. As many other well-studied bursts, the overall temporal and spectral behavior of the afterglow is hard to explain in the frame of the standard fireball model. 

\acknowledgments
FY gratefully acknowledges useful discussions with S.~B.~Pandey. This work made use of data supplied by the UK {\em Swift} Science Data Centre at the University of Leicester. FY is supported by the NASA {\em Swift} Guest Investigator grants NNX-08AN25G. JLR acknowledges support for this work from NASA contract NAS5-00136. RW and PTO gratefully acknowledge funding from the STFC for {\em Swift} at the University of Leicester. TK acknowledges support by the DFG cluster of excellence `Origin and Structure of the Universe'. ROTSE-III has been supported by NASA grant NNX-08AV63G, NSF grant PHY-0801007, the Australian Research Council, the University of New South Wales, the University of Texas, and the University of Michigan. Special thanks to the H.E.S.S. staff, especially Toni Hanke. Part of the funding for GROND (both hardware as well as personnel) was generously granted from the Leibniz-Prize (DFG grant HA 1850/28-1) to Prof. G. Hasinger (IPP).


\begin{thebibliography}

\bibitem[{{Akerlof} {et~al.}(1999){Akerlof}, {Balsano}, {Barthelemy}, {Bloch},
  {Butterworth}, {Casperson}, {Cline}, {Fletcher}, {Frontera}, {Gisler},
  {Heise}, {Hills}, {Kehoe}, {Lee}, {Marshall}, {McKay}, {Miller}, {Piro},
  {Priedhorsky}, {Szymanski}, \& {Wren}}]{abbbb99}
 {Akerlof}, C., et~al. 1999, \nat, 398, 400

\bibitem[{{Akerlof} {et~al.}(2003){Akerlof}, {Kehoe}, {McKay}, {Rykoff},
  {Smith}, {Casperson}, {McGowan}, {Vestrand}, {Wozniak}, {Wren}, {Ashley},
  {Phillips}, {Marshall}, {Epps}, \& {Schier}}]{akmrs03}
 {Akerlof}, C.~W., et~al. 2003, \pasp, 115, 132

\bibitem[{{Band} {et~al.}(1993){Band}, {Matteson}, {Ford}, {Schaefer},
  {Palmer}, {Teegarden}, {Cline}, {Briggs}, {Paciesas}, {Pendleton}, {Fishman},
  {Kouveliotou}, {Meegan}, {Wilson}, \& {Lestrade}}]{bmfsp93}
 {Band}, D., et~al. 1993, \apj, 413, 281

\bibitem[{{Bertin} \& {Arnouts}(1996)}]{ba96}
{Bertin}, E. \& {Arnouts}, S. 1996, \aaps, 117, 393

\bibitem[{{Bloom} {et~al.}(2003){Bloom}, {Frail}, \& {Kulkarni}}]{bfk03}
{Bloom}, J.~S., {Frail}, D.~A., \& {Kulkarni}, S.~R. 2003, \apj, 594, 674

\bibitem[{{Burrows} {et~al.}(2005){Burrows}, {Romano}, {Falcone}, {Kobayashi},
  {Zhang}, {Moretti}, {O'Brien}, {Goad}, {Campana}, {Page}, {Angelini},
  {Barthelmy}, {Beardmore}, {Capalbi}, {Chincarini}, {Cummings}, {Cusumano},
  {Fox}, {Giommi}, {Hill}, {Kennea}, {Krimm}, {Mangano}, {Marshall},
  {M{\'e}sz{\'a}ros}, {Morris}, {Nousek}, {Osborne}, {Pagani}, {Perri},
  {Tagliaferri}, {Wells}, {Woosley}, \& {Gehrels}}]{brfkz05}
 {Burrows}, D.~N., et~al. 2005, Science, 309, 1833

\bibitem[{{Chevalier} \& {Li}(2000)}]{cl00}
{Chevalier}, R.~A. \& {Li}, Z.-Y. 2000, \apj, 536, 195

\bibitem[{{Chincarini} {et~al.}(2007){Chincarini}, {Moretti}, {Romano},
  {Falcone}, {Morris}, {Racusin}, {Campana}, {Covino}, {Guidorzi},
  {Tagliaferri}, {Burrows}, {Pagani}, {Stroh}, {Grupe}, {Capalbi}, {Cusumano},
  {Gehrels}, {Giommi}, {La Parola}, {Mangano}, {Mineo}, {Nousek}, {O'Brien},
  {Page}, {Perri}, {Troja}, {Willingale}, \& {Zhang}}]{cmrfm07}
 {Chincarini}, G., et~al. 2007, \apj, 671, 1903

\bibitem[{{Cucchiara} {et~al.}(2008{\natexlab{a}}){Cucchiara}, {Fox}, {Cenko},
  \& {Berger}}]{cfcb08g2}
{Cucchiara}, A., {Fox}, D.~B., {Cenko}, S.~B., \& {Berger}, E.
  2008{\natexlab{a}}, GRB Coordinates Network, 8372

\bibitem[{{Cucchiara} {et~al.}(2008{\natexlab{b}}){Cucchiara}, {Fox}, {Cenko},
  \& {Berger}}]{cfcb08g1}
{Cucchiara}, A., {Fox}, D.~B., {Cenko}, S.~B., \& {Berger}, E.
  2008{\natexlab{b}}, GRB Coordinates Network, 8346

\bibitem[{{D'Avanzo} {et~al.}(2008){D'Avanzo}, {D'Elia}, \& {Covino}}]{ddc08g2}
{D'Avanzo}, P., {D'Elia}, V., \& {Covino}, S. 2008, GRB Coordinates Network,
  8350

\bibitem[{{Evans} {et~al.}(2009){Evans}, {Beardmore}, {Page}, {Osborne},
  {O'Brien}, {Willingale}, {Starling}, {Burrows}, {Godet}, {Vetere}, {Racusin},
  {Goad}, {Wiersema}, {Angelini}, {Capalbi}, {Chincarini}, {Gehrels}, {Kennea},
  {Margutti}, {Morris}, {Mountford}, {Pagani}, {Perri}, {Romano}, \&
  {Tanvir}}]{ebpoo09}
 {Evans}, P.~A., et~al. 2009, \mnras, 397, 1177

\bibitem[{{Evans} {et~al.}(2007){Evans}, {Beardmore}, {Page}, {Tyler},
  {Osborne}, {Goad}, {O'Brien}, {Vetere}, {Racusin}, {Morris}, {Burrows},
  {Capalbi}, {Perri}, {Gehrels}, \& {Romano}}]{ebpto07}
 {Evans}, P.~A., et~al. 2007, \aap, 469, 379

\bibitem[{{Falcone} {et~al.}(2006){Falcone}, {Burrows}, {Lazzati}, {Campana},
  {Kobayashi}, {Zhang}, {M{\'e}sz{\'a}ros}, {Page}, {Kennea}, {Romano},
  {Pagani}, {Angelini}, {Beardmore}, {Capalbi}, {Chincarini}, {Cusumano},
  {Giommi}, {Goad}, {Godet}, {Grupe}, {Hill}, {La Parola}, {Mangano},
  {Moretti}, {Nousek}, {O'Brien}, {Osborne}, {Perri}, {Tagliaferri}, {Wells},
  \& {Gehrels}}]{fblck06}
 {Falcone}, A.~D., et~al. 2006, \apj, 641, 1010

\bibitem[{{Falcone} {et~al.}(2007){Falcone}, {Morris}, {Racusin}, {Chincarini},
  {Moretti}, {Romano}, {Burrows}, {Pagani}, {Stroh}, {Grupe}, {Campana},
  {Covino}, {Tagliaferri}, {Willingale}, \& {Gehrels}}]{fmrcm07}
 {Falcone}, A.~D., et~al. 2007, \apj, 671, 1921

\bibitem[{{Gehrels} {et~al.}(2004){Gehrels}, {Chincarini}, {Giommi}, {Mason},
  {Nousek}, {Wells}, {White}, {Barthelmy}, {Burrows}, {Cominsky}, {Hurley},
  {Marshall}, {M{\' e}sz{\' a}ros}, {Roming}, {Angelini}, {Barbier}, {Belloni},
  {Campana}, {Caraveo}, {Chester}, {Citterio}, {Cline}, {Cropper}, {Cummings},
  {Dean}, {Feigelson}, {Fenimore}, {Frail}, {Fruchter}, {Garmire}, {Gendreau},
  {Ghisellini}, {Greiner}, {Hill}, {Hunsberger}, {Krimm}, {Kulkarni}, {Kumar},
  {Lebrun}, {Lloyd-Ronning}, {Markwardt}, {Mattson}, {Mushotzky}, {Norris},
  {Osborne}, {Paczynski}, {Palmer}, {Park}, {Parsons}, {Paul}, {Rees},
  {Reynolds}, {Rhoads}, {Sasseen}, {Schaefer}, {Short}, {Smale}, {Smith},
  {Stella}, {Tagliaferri}, {Takahashi}, {Tashiro}, {Townsley}, {Tueller},
  {Turner}, {Vietri}, {Voges}, {Ward}, {Willingale}, {Zerbi}, \&
  {Zhang}}]{gcgmn04}
 {Gehrels}, N., et~al. 2004, \apj, 611, 1005

\bibitem[{{Granot} {et~al.}(2003){Granot}, {Nakar}, \& {Piran}}]{gnp03}
{Granot}, J., {Nakar}, E., \& {Piran}, T. 2003, \nat, 426, 138

\bibitem[{{Granot} \& {Sari}(2002)}]{gs02}
{Granot}, J. \& {Sari}, R. 2002, \apj, 568, 820

\bibitem[{{Greiner} {et~al.}(2008){Greiner}, {Bornemann}, {Clemens}, {Deuter},
  {Hasinger}, {Honsberg}, {Huber}, {Huber}, {Krauss}, {Kr{\"u}hler},
  {K{\"u}pc{\"u} Yolda{\c s}}, {Mayer-Hasselwander}, {Mican}, {Primak},
  {Schrey}, {Steiner}, {Szokoly}, {Th{\"o}ne}, {Yolda{\c s}}, {Klose}, {Laux},
  \& {Winkler}}]{gbcdh08}
 {Greiner}, J., et~al. 2008, \pasp, 120, 405

\bibitem[{{Greiner} {et~al.}(2009){Greiner}, {Kr{\"u}hler}, {McBreen},
  {Ajello}, {Giannios}, {Schwarz}, {Savaglio}, {Yolda{\c s}}, {Clemens},
  {Stefanescu}, {Sala}, {Bertoldi}, {Szokoly}, \& {Klose}}]{gkmag09}
 {Greiner}, J., et~al. 2009, \apj, 693, 1912

\bibitem[{{Guidorzi} {et~al.}(2009){Guidorzi}, {Clemens}, {Kobayashi},
  {Granot}, {Melandri}, {D'Avanzo}, {Kuin}, {Klotz}, {Fynbo}, {Covino},
  {Greiner}, {Malesani}, {Mao}, {Mundell}, {Steele}, {Jakobsson}, {Margutti},
  {Bersier}, {Campana}, {Chincarini}, {D'Elia}, {Fugazza}, {Genet}, {Gomboc},
  {Kr{\"u}hler}, {K{\"u}pc{\"u} Yolda{\c s}}, {Moretti}, {Mottram}, {O'Brien},
  {Smith}, {Szokoly}, {Tagliaferri}, {Tanvir}, \& {Gehrels}}]{gckgm09}
 {Guidorzi}, C., et~al. 2009, \aap, 499, 439

\bibitem[{{Jin} \& {Fan}(2007)}]{jf07}
{Jin}, Z.~P. \& {Fan}, Y.~Z. 2007, \mnras, 378, 1043

\bibitem[{{J{\'o}hannesson} {et~al.}(2006){J{\'o}hannesson}, {Bj{\"o}rnsson},
  \& {Gudmundsson}}]{jbg06}
{J{\'o}hannesson}, G., {Bj{\"o}rnsson}, G., \& {Gudmundsson}, E.~H. 2006, \apj,
  647, 1238

\bibitem[{{Kalberla} {et~al.}(2005){Kalberla}, {Burton}, {Hartmann}, {Arnal},
  {Bajaja}, {Morras}, \& {P{\"o}ppel}}]{kbhab05}
{Kalberla}, P.~M.~W., {Burton}, W.~B., {Hartmann}, D., {Arnal}, E.~M.,
  {Bajaja}, E., {Morras}, R., \& {P{\"o}ppel}, W.~G.~L. 2005, \aap, 440, 775

\bibitem[{{Klotz} {et~al.}(2009){Klotz}, {Bo{\"e}r}, {Atteia}, \&
  {Gendre}}]{kbag09}
{Klotz}, A., {Bo{\"e}r}, M., {Atteia}, J.~L., \& {Gendre}, B. 2009, \aj, 137,
  4100

\bibitem[{{Kobayashi}(2000)}]{k00}
{Kobayashi}, S. 2000, \apj, 545, 807

\bibitem[{{Kr{\"u}hler} {et~al.}(2009){Kr{\"u}hler}, {Greiner}, {McBreen},
  {Klose}, {Rossi}, {Afonso}, {Clemens}, {Filgas}, {Yolda{\c s}}, {Szokoly}, \&
  {Yolda{\c s}}}]{kgmkr09}
 {Kr{\"u}hler}, T., et~al. 2009, \apj, 697, 758

\bibitem[{{Kumar} \& {Panaitescu}(2000)}]{kpa00}
{Kumar}, P. \& {Panaitescu}, A. 2000, \apjl, 541, L51

\bibitem[{{Kumar} \& {Panaitescu}(2003)}]{kp03}
{Kumar}, P. \& {Panaitescu}, A. 2003, \mnras, 346, 905

\bibitem[{{Kumar} \& {Piran}(2000)}]{kp00}
{Kumar}, P. \& {Piran}, T. 2000, \apj, 532, 286

\bibitem[{{Lazzati} \& {Begelman}(2006)}]{lb06}
{Lazzati}, D. \& {Begelman}, M.~C. 2006, \apj, 641, 972

\bibitem[{{Liang} {et~al.}(2008){Liang}, {Racusin}, {Zhang}, {Zhang}, \&
  {Burrows}}]{lrzzb08}
{Liang}, E.-W., {Racusin}, J.~L., {Zhang}, B., {Zhang}, B.-B., \& {Burrows},
  D.~N. 2008, \apj, 675, 528

\bibitem[{{Liang} {et~al.}(2006){Liang}, {Zhang}, {O'Brien}, {Willingale},
  {Angelini}, {Burrows}, {Campana}, {Chincarini}, {Falcone}, {Gehrels}, {Goad},
  {Grupe}, {Kobayashi}, {M{\'e}sz{\'a}ros}, {Nousek}, {Osborne}, {Page}, \&
  {Tagliaferri}}]{lzowa06}
 {Liang}, E.~W., et~al. 2006, \apj, 646, 351

\bibitem[{{Liang} {et~al.}(2007){Liang}, {Zhang}, \& {Zhang}}]{lzz07}
{Liang}, E.-W., {Zhang}, B.-B., \& {Zhang}, B. 2007, \apj, 670, 565

\bibitem[{{Meszaros}(2006)}]{mesza06}
{Meszaros}, P. 2006, Reports on Progress in Physics, 69, 2259

\bibitem[{{Molinari} {et~al.}(2007){Molinari}, {Vergani}, {Malesani}, {Covino},
  {D'Avanzo}, {Chincarini}, {Zerbi}, {Antonelli}, {Conconi}, {Testa}, {Tosti},
  {Vitali}, {D'Alessio}, {Malaspina}, {Nicastro}, {Palazzi}, {Guetta},
  {Campana}, {Goldoni}, {Masetti}, {Meurs}, {Monfardini}, {Norci}, {Pian},
  {Piranomonte}, {Rizzuto}, {Stefanon}, {Stella}, {Tagliaferri}, {Ward},
  {Ihle}, {Gonzalez}, {Pizarro}, {Sinclaire}, \& {Valenzuela}}]{mvmcd07}
 {Molinari}, E., et~al. 2007, \aap, 469, L13

\bibitem[{{Nousek} {et~al.}(2006){Nousek}, {Kouveliotou}, {Grupe}, {Page},
  {Granot}, {Ramirez-Ruiz}, {Patel}, {Burrows}, {Mangano}, {Barthelmy},
  {Beardmore}, {Campana}, {Capalbi}, {Chincarini}, {Cusumano}, {Falcone},
  {Gehrels}, {Giommi}, {Goad}, {Godet}, {Hurkett}, {Kennea}, {Moretti},
  {O'Brien}, {Osborne}, {Romano}, {Tagliaferri}, \& {Wells}}]{nkgpg06}
 {Nousek}, J.~A., et~al. 2006, \apj, 642, 389

\bibitem[{{Oates} {et~al.}(2009){Oates}, {Page}, {Schady}, {de Pasquale},
  {Koch}, {Breeveld}, {Brown}, {Chester}, {Holland}, {Hoversten}, {Kuin},
  {Marshall}, {Roming}, {Still}, {vanden Berk}, {Zane}, \& {Nousek}}]{opsdk09}
 {Oates}, S.~R., et~al. 2009, \mnras, 395, 490

\bibitem[{{O'Brien} {et~al.}(2006){O'Brien}, {Willingale}, {Osborne}, {Goad},
  {Page}, {Vaughan}, {Rol}, {Beardmore}, {Godet}, {Hurkett}, {Wells}, {Zhang},
  {Kobayashi}, {Burrows}, {Nousek}, {Kennea}, {Falcone}, {Grupe}, {Gehrels},
  {Barthelmy}, {Cannizzo}, {Cummings}, {Hill}, {Krimm}, {Chincarini},
  {Tagliaferri}, {Campana}, {Moretti}, {Giommi}, {Perri}, {Mangano}, \&
  {LaParola}}]{owogp06}
 {O'Brien}, P.~T., et~al. 2006, \apj, 647, 1213

\bibitem[{{Page} {et~al.}(2009){Page}, {Willingale}, {Bissaldi}, {Postigo},
  {Holland}, {McBreen}, {O'Brien}, {Osborne}, {Prochaska}, {Rol}, {Rykoff},
  {Starling}, {Tanvir}, {van der Horst}, {Wiersema}, {Zhang}, {Aceituno},
  {Akerlof}, {Beardmore}, {Briggs}, {Burrows}, {Castro-Tirado}, {Connaughton},
  {Evans}, {Fynbo}, {Gehrels}, {Guidorzi}, {Howard}, {Kennea}, {Kouveliotou},
  {Pagani}, {Preece}, {Perley}, {Steele}, \& {Yuan}}]{pwbdh09}
 {Page}, K.~L., et~al. 2009, \mnras, 400, 134

\bibitem[{{Page} {et~al.}(2007){Page}, {Willingale}, {Osborne}, {Zhang},
  {Godet}, {Marshall}, {Melandri}, {Norris}, {O'Brien}, {Pal'shin}, {Rol},
  {Romano}, {Starling}, {Schady}, {Yost}, {Barthelmy}, {Beardmore}, {Cusumano},
  {Burrows}, {De Pasquale}, {Ehle}, {Evans}, {Gehrels}, {Goad}, {Golenetskii},
  {Guidorzi}, {Mundell}, {Page}, {Ricker}, {Sakamoto}, {Schaefer},
  {Stamatikos}, {Troja}, {Ulanov}, {Yuan}, \& {Ziaeepour}}]{pwozg07}
 {Page}, K.~L., et~al. 2007, \apj, 663, 1125

\bibitem[{{Panaitescu}(2005)}]{pana05}
{Panaitescu}, A. 2005, \mnras, 362, 921

\bibitem[{{Panaitescu} \& {Vestrand}(2008)}]{pv08}
{Panaitescu}, A. \& {Vestrand}, W.~T. 2008, \mnras, 387, 497

\bibitem[{{Quimby} {et~al.}(2006){Quimby}, {Rykoff}, {Yost}, {Aharonian},
  {Akerlof}, {Alatalo}, {Ashley}, {G{\"o}{\u g}{\"u}{\c s}}, {G{\"u}ver},
  {Horns}, {Kehoe}, {K{$\iota$}z{$\iota$}lo{\u g}lu}, {Mckay}, {{\"O}zel},
  {Phillips}, {Schaefer}, {Smith}, {Swan}, {Vestrand}, {Wheeler}, \&
  {Wren}}]{qryaa06}
 {Quimby}, R.~M., et~al. 2006, \apj, 640, 402

\bibitem[{{Racusin} {et~al.}(2008{\natexlab{a}}){Racusin}, {Baumgartner},
  {Brown}, {Burrows}, {Evans}, {Grupe}, {Holland}, {Hoversten}, {Kennea}, {La
  Parola}, {Marshall}, {O'Brien}, {Osborne}, {Parsons}, {Stamatikos},
  {Starling}, {Tagliaferri}, {Ukwatta}, \& {Vetere}}]{rbbbe08}
 {Racusin}, J.~L., et~al. 2008{\natexlab{a}}, GRB Coordinates Network, 8344

\bibitem[{{Racusin} {et~al.}(2008{\natexlab{b}}){Racusin}, {Karpov},
  {Sokolowski}, {Granot}, {Wu}, {Pal'Shin}, {Covino}, {van der Horst}, {Oates},
  {Schady}, {Smith}, {Cummings}, {Starling}, {Piotrowski}, {Zhang}, {Evans},
  {Holland}, {Malek}, {Page}, {Vetere}, {Margutti}, {Guidorzi}, {Kamble},
  {Curran}, {Beardmore}, {Kouveliotou}, {Mankiewicz}, {Melandri}, {O'Brien},
  {Page}, {Piran}, {Tanvir}, {Wrochna}, {Aptekar}, {Barthelmy}, {Bartolini},
  {Beskin}, {Bondar}, {Bremer}, {Campana}, {Castro-Tirado}, {Cucchiara},
  {Cwiok}, {D'Avanzo}, {D'Elia}, {Della Valle}, {de Ugarte Postigo}, {Dominik},
  {Falcone}, {Fiore}, {Fox}, {Frederiks}, {Fruchter}, {Fugazza}, {Garrett},
  {Gehrels}, {Golenetskii}, {Gomboc}, {Gorosabel}, {Greco}, {Guarnieri},
  {Immler}, {Jelinek}, {Kasprowicz}, {La Parola}, {Levan}, {Mangano}, {Mazets},
  {Molinari}, {Moretti}, {Nawrocki}, {Oleynik}, {Osborne}, {Pagani}, {Pandey},
  {Paragi}, {Perri}, {Piccioni}, {Ramirez-Ruiz}, {Roming}, {Steele}, {Strom},
  {Testa}, {Tosti}, {Ulanov}, {Wiersema}, {Wijers}, {Winters}, {Zarnecki},
  {Zerbi}, {M{\'e}sz{\'a}ros}, {Chincarini}, \& {Burrows}}]{rksgw08}
 {Racusin}, J.~L., et~al. 2008{\natexlab{b}}, \nat, 455, 183

\bibitem[{{Racusin} {et~al.}(2009){Racusin}, {Liang}, {Burrows}, {Falcone},
  {Sakamoto}, {Zhang}, {Zhang}, {Evans}, \& {Osborne}}]{rlbfs09}
{Racusin}, J.~L., {Liang}, E.~W., {Burrows}, D.~N., {Falcone}, A., {Sakamoto},
  T., {Zhang}, B.~B., {Zhang}, B., {Evans}, P., \& {Osborne}, J. 2009, \apj,
  698, 43

\bibitem[{{Rees} \& {Meszaros}(1998)}]{rm98}
{Rees}, M.~J. \& {Meszaros}, P. 1998, \apjl, 496, L1

\bibitem[{{Romano} {et~al.}(2006){Romano}, {Moretti}, {Banat}, {Burrows},
  {Campana}, {Chincarini}, {Covino}, {Malesani}, {Tagliaferri}, {Kobayashi},
  {Zhang}, {Falcone}, {Angelini}, {Barthelmy}, {Beardmore}, {Capalbi},
  {Cusumano}, {Giommi}, {Goad}, {Godet}, {Grupe}, {Hill}, {Kennea}, {La
  Parola}, {Mangano}, {M{\'e}sz{\'a}ros}, {Morris}, {Nousek}, {O'Brien},
  {Osborne}, {Parsons}, {Perri}, {Pagani}, {Page}, {Wells}, \&
  {Gehrels}}]{rmbbc06}
 {Romano}, P., et~al. 2006, \aap, 450, 59

\bibitem[{{Rykoff}(2008)}]{r08}
{Rykoff}, E.~S. 2008, GRB Coordinates Network, 8343

\bibitem[{{Rykoff} {et~al.}(2009){Rykoff}, {Aharonian}, {Akerlof}, {Ashley},
  {Barthelmy}, {Flewelling}, {Gehrels}, {G{\"o}{\v g}{\"u}{\c s}}, {G{\"u}ver},
  {Kizilo{\v g}lu}, {Krimm}, {McKay}, {{\"O}zel}, {Phillips}, {Quimby},
  {Rowell}, {Rujopakarn}, {Schaefer}, {Smith}, {Vestrand}, {Wheeler}, {Wren},
  {Yuan}, \& {Yost}}]{eaaab09}
 {Rykoff}, E.~S., et~al. 2009, \apj, 702, 489

\bibitem[{{Rykoff} {et~al.}(2006){Rykoff}, {Mangano}, {Yost}, {Sari},
  {Aharonian}, {Akerlof}, {Ashley}, {Barthelmy}, {Burrows}, {Gehrels},
  {G{\"o}{\v g}{\"u}{\c s}}, {G{\"u}ver}, {Horns}, {K{\i}z{\i}lo{\v g}lu},
  {Krimm}, {McKay}, {{\"O}zel}, {Phillips}, {Quimby}, {Rowell}, {Rujopakarn},
  {Schaefer}, {Smith}, {Swan}, {Vestrand}, {Wheeler}, {Wren}, \&
  {Yuan}}]{rmysa06}
 {Rykoff}, E.~S., et~al. 2006, \apjl, 638, L5

\bibitem[{{Rykoff} {et~al.}(2005){Rykoff}, {Yost}, {Krimm}, {Aharonian},
  {Akerlof}, {Alatalo}, {Ashley}, {Barthelmy}, {Gehrels},
  {G{\"o}{\u{g}}{\"u}{\c s}}, {G{\"u}ver}, {Horns}, {K{\i}z{\i}l{\v o}lu},
  {McKay}, {{\"O}zel}, {Phillips}, {Quimby}, {Rujopakarn}, {Schaefer}, {Smith},
  {Swan}, {Vestrand}, {Wheeler}, \& {Wren}}]{rykaa05}
 {Rykoff}, E.~S., et~al. 2005, \apjl, 631, L121

\bibitem[{{Sakamoto} {et~al.}(2009){Sakamoto}, {Sato}, {Barbier}, {Barthelmy},
  {Cummings}, {Fenimore}, {Gehrels}, {Hullinger}, {Krimm}, {Lamb}, {Markwardt},
  {Palmer}, {Parsons}, {Stamatikos}, {Tueller}, \& {Ukwatta}}]{ssbbc09}
 {Sakamoto}, T., et~al. 2009, \apj, 693, 922

\bibitem[{{Sari} \& {M{\'e}sz{\'a}ros}(2000)}]{sm00}
{Sari}, R. \& {M{\'e}sz{\'a}ros}, P. 2000, \apjl, 535, L33

\bibitem[{{Sari} \& {Piran}(1999)}]{sp99}
{Sari}, R. \& {Piran}, T. 1999, \apj, 520, 641

\bibitem[{{Sari} {et~al.}(1999){Sari}, {Piran}, \& {Halpern}}]{sph99}
{Sari}, R., {Piran}, T., \& {Halpern}, J.~P. 1999, \apjl, 519, L17

\bibitem[{{Scargle}(1998)}]{s98}
{Scargle}, J.~D. 1998, \apj, 504, 405

\bibitem[{{Schlegel} {et~al.}(1998){Schlegel}, {Finkbeiner}, \&
  {Davis}}]{sfd98}
{Schlegel}, D.~J., {Finkbeiner}, D.~P., \& {Davis}, M. 1998, \apj, 500, 525

\bibitem[{{Skrutskie} {et~al.}(2006){Skrutskie}, {Cutri}, {Stiening},
  {Weinberg}, {Schneider}, {Carpenter}, {Beichman}, {Capps}, {Chester},
  {Elias}, {Huchra}, {Liebert}, {Lonsdale}, {Monet}, {Price}, {Seitzer},
  {Jarrett}, {Kirkpatrick}, {Gizis}, {Howard}, {Evans}, {Fowler}, {Fullmer},
  {Hurt}, {Light}, {Kopan}, {Marsh}, {McCallon}, {Tam}, {Van Dyk}, \&
  {Wheelock}}]{scsws06}
 {Skrutskie}, M.~F., et~al. 2006, \aj, 131, 1163

\bibitem[{{Smith} {et~al.}(2002){Smith}, {Tucker}, {Kent}, {Richmond},
  {Fukugita}, {Ichikawa}, {Ichikawa}, {Jorgensen}, {Uomoto}, {Gunn}, {Hamabe},
  {Watanabe}, {Tolea}, {Henden}, {Annis}, {Pier}, {McKay}, {Brinkmann}, {Chen},
  {Holtzman}, {Shimasaku}, \& {York}}]{stkrf02}
 {Smith}, J.~A., et~al. 2002, \aj, 123, 2121

\bibitem[{{Stetson}(1987)}]{stetson87}
{Stetson}, P.~B. 1987, \pasp, 99, 191

\bibitem[{{Tagliaferri} {et~al.}(2005){Tagliaferri}, {Goad}, {Chincarini},
  {Moretti}, {Campana}, {Burrows}, {Perri}, {Barthelmy}, {Gehrels}, {Krimm},
  {Sakamoto}, {Kumar}, {M{\'e}sz{\'a}ros}, {Kobayashi}, {Zhang}, {Angelini},
  {Banat}, {Beardmore}, {Capalbi}, {Covino}, {Cusumano}, {Giommi}, {Godet},
  {Hill}, {Kennea}, {Mangano}, {Morris}, {Nousek}, {O'Brien}, {Osborne},
  {Pagani}, {Page}, {Romano}, {Stella}, \& {Wells}}]{tgcmc05}
 {Tagliaferri}, G., et~al. 2005, \nat, 436, 985

\bibitem[{{Tody}(1993)}]{tody93}
{Tody}, D. 1993, in Astronomical Society of the Pacific Conference Series,
  Vol.~52, Astronomical Data Analysis Software and Systems II, ed. R.~J.
  {Hanisch}, R.~J.~V. {Brissenden}, \& J.~{Barnes}, 173

\bibitem[{{Vestrand} {et~al.}(2005){Vestrand}, {Wozniak}, {Wren}, {Fenimore},
  {Sakamoto}, {White}, {Casperson}, {Davis}, {Evans}, {Galassi}, {McGowan},
  {Schier}, {Asa}, {Barthelmy}, {Cummings}, {Gehrels}, {Hullinger}, {Krimm},
  {Markwardt}, {McLean}, {Palmer}, {Parsons}, \& {Tueller}}]{vwwfs05}
 {Vestrand}, W.~T., et~al. 2005, \nat, 435, 178

\bibitem[{{Vestrand} {et~al.}(2006){Vestrand}, {Wren}, {Wozniak}, {Aptekar},
  {Golentskii}, {Pal'Shin}, {Sakamoto}, {White}, {Evans}, {Casperson}, \&
  {Fenimore}}]{vwwag06}
 {Vestrand}, W.~T., et~al. 2006, \nat, 442, 172

\bibitem[{{Willingale} {et~al.}(2007){Willingale}, {O'Brien}, {Osborne},
  {Godet}, {Page}, {Goad}, {Burrows}, {Zhang}, {Rol}, {Gehrels}, \&
  {Chincarini}}]{woogp07}
 {Willingale}, R., et~al. 2007, \apj, 662, 1093

\bibitem[{{Yuan} \& {Akerlof}(2008)}]{ya08}
{Yuan}, F. \& {Akerlof}, C.~W. 2008, \apj, 677, 808

\bibitem[{{Yuan} {et~al.}(2008){Yuan}, {Rykoff}, {Schaefer}, {Rujopakarn},
  {G{\"u}ver}, {Aharonian}, {Akerlof}, {Ashley}, {Barthelmy}, {Gehrels},
  {G{\"o}{\v g}{\"u}{\c s}}, {Horns}, {Kizilo{\v g}lu}, {Krimm}, {McKay},
  {{\"O}zel}, {Phillips}, {Quimby}, {Rowell}, {Swan}, {Vestrand}, {Wheeler}, \&
  {Wren}}]{yrsrg08}
 {Yuan}, F., et~al. 2008, in American Institute of Physics Conference Series,
  Vol. 1065, American Institute of Physics Conference Series, ed. Y.-F.
  {Huang}, Z.-G. {Dai}, \& B.~{Zhang}, 103--106

\bibitem[{{Zhang} {et~al.}(2006){Zhang}, {Fan}, {Dyks}, {Kobayashi},
  {M{\'e}sz{\'a}ros}, {Burrows}, {Nousek}, \& {Gehrels}}]{zfdkm06}
{Zhang}, B., {Fan}, Y.~Z., {Dyks}, J., {Kobayashi}, S., {M{\'e}sz{\'a}ros}, P.,
  {Burrows}, D.~N., {Nousek}, J.~A., \& {Gehrels}, N. 2006, \apj, 642, 354

\bibitem[{{Zhang} {et~al.}(2003){Zhang}, {Kobayashi}, \&
  {M{\'e}sz{\'a}ros}}]{zkm03}
{Zhang}, B., {Kobayashi}, S., \& {M{\'e}sz{\'a}ros}, P. 2003, \apj, 595, 950

\bibitem[{{Zhang} {et~al.}(2007){Zhang}, {Liang}, \& {Zhang}}]{zlz07}
{Zhang}, B.-B., {Liang}, E.-W., \& {Zhang}, B. 2007, \apj, 666, 1002

\end{thebibliography}
\end{document}